# Abnormality of the placental vasculature affects placental thickness.

**Michael Yampolsky[1], Carolyn M. Salafia[2,3], Oleksandr Shlakhter[4],**

**Danielle Haas[2], Barbara Eucker[5], John Thorp[5]**

[1] Department of Mathematics, University of Toronto, 40 St. George St, Toronto, Ontario, Canada, M5S2E4

[2] Placental Analytics, LLC, Larchmont, NY, USA

[3] Department of Obstetrics and Gynecology and Pediatrics, New York Methodist Hospital, Brooklyn, NY.

[4] Rotman School of Management, University of Toronto, 105 St. George Street, Toronto, ON, M5S 3E6, Canada

[5] Department of Obstetrics and Gynecology, University of North Carolina, Chapel Hill, Chapel Hill, North Carolina.

**Corresponding author:**
Michael Yampolsky, PhD
Department of Mathematics
University of Toronto
40 St. George Street, Toronto, Canada M5S2E4
Email: yampolsky.michael@gmail.com

**Abstract.**

***Background.*** Our empirical modeling suggests that deformation of placental vascular growth is associated with abnormal placental chorionic surface shape. Altered chorionic surface shape is associated with lowered placental functional efficiency. We hypothesize that placentas with deformed chorionic surface vascular trees and reduced functional efficiency also have irregular vascular arborization that will be reflected in increased variability of placental thickness and a lower mean thickness.

***Materials and Methods.*** The sample was drawn from the *Pregnancy, Infection, and Nutrition Study*, a cohort in which 94.6 percent of women consented to placental examination. Of these, 87% had placentas collected and photographed for chorionic surface analysis. Of the 1023 delivered at term, intact placentas were fixed, sliced and photographed in 587 (57%) cases.. Fourier analysis was used to quantify the non-centrality of the umbilical cord insertion, and regularly irregular chorionic surface shapes. To quantify the disk thickness, one slice of the placenta at the mid-section was used to calculate the mean and variance of placental thickness. Functional efficiency was quantified as the scaling exponent $\beta$=log(Placental Weight)/log(Birth Weight, normally 0.75)with higher $\beta$ corresponding to lower functional efficiency. Spearman's correlations considered $p<0.05$ significant.

***Results.*** Non-centrality of the umbilical cord insertion was strongly and significantly positively correlated with disk thickness of the (Spearman's $\rho=0.128$, $p=0.002$). Deformed shape was strongly and significantly associated with lower overall thickness and higher variability of thickness with $\rho$ between -0.173 and -0.254 ($p<0.001$) . Both lower mean thickness and high variability of thickness are strongly correlated with higher $\beta$ (reduced placental efficiency) ($p<0.001$ and $p=0.038$ respectively). Greater thickness variability is correlated with higher $\beta$ independent of the other placental shape variables $p=0.004$.

***Conclusions.*** Our findings confirm the predictions of our empirical modeling. A regularly deformed placental surface is associated with non-uniform placental thickness and lower average thickness. We speculate that lower mean thickness or high variability of thickness are caused by variable villous arborization due to abnormal placental vascular growth. Supporting this, placentas with lower average thickness or higher variability of thickness are less functionally efficient. Variability of placental disk thickness, even when restricted to analysis of a single random diameter, is particularly useful as an independent indicator of lowered placental efficiency, and thus abnormal placental vascular growth.

Keywords: Placenta, placental weight, placental thickness, birth weight, placental function, chorionic plate

## Introduction.

In our previous work [1,2] we described how a deformation of the placental vascular growth is reflected in an abnormal placental surface shape and reduced placental functional efficiency. Commonly, when the angiogenesis of the vascular tree is impacted in our model [1], the growth of the branches of the vascular tree is depressed below a certain level. This results in a model tree with gaps in the vascular coverage concentrated in sectors between larger vascular branches. The placental surface will typically be deformed into a commonly observed "regularly-irregular" multilobate or star-like shape. We have conjectured that the gaps in the vascular penetration by a deformed vascular tree will also lead to an increased variability of placental cross-sectional thickness and lowered overall thickness. Reduced vascular arborization, is in turn associated with lower placental functional efficiency, as shown in [2].

In the present work we test these hypotheses by analyzing the relation of the placental thickness, measured in a single randomly oriented cross-section through the center of the placental surface, with the placental shape. We use Fourier analysis to quantify the shape of the chorionic plate, since it can quantify the typical "regularly-irregular" placental shapes described in [1].

If variable thickness does in fact reflect variable villous arborization and, by extension, vascularity, then variability of placental thickness would also be expected to have an effect on the placental functional efficiency, quantified as the fetoplacental scaling exponent $\beta$ [3,4].

To test our methods of measuring placental thickness, we can apply them to study the impact of non-centrality of umbilical cord insertion on thickness. As shown in [2], non-central insertion is not associated with a change in placental surface shape, and is significantly correlated with a larger placental weight, and lower functional efficiency (smaller baby for a given placental weight). We thus expect to find that a non-centrality of cord insertion produces a measurable increase in placental thickness, which for the same surface area would result in a heavier placenta.

## Materials and Methods.

### *Placental Cohort*

The *Pregnancy, Infection, and Nutrition Study* is a cohort study of pregnant women recruited at mid pregnancy from an academic health center in central North Carolina. Our study population and recruitment techniques are described in detail elsewhere [5]. Beginning in March 2002, all women recruited into the *Pregnancy, Infection, and Nutrition Study* were requested to consent to a detailed placental examination. As of October 1, 2004, 94.6 percent of women consented to such examination. Of those women who consented, 87 percent had placentas collected and photographed for image analysis. Of the 1,225 consecutive placentas collected, 1023 were delivered at term. Placental gross examinations, histology review, and image analyses were performed at *EarlyPath Clinical and Research Diagnostics*, a New York State-licensed histopathology facility under the direct supervision of Dr. Salafia. The Institutional Review Board of the University of North Carolina at Chapel Hill approved this protocol.

The fetal surface of the placenta was wiped dry and placed on a clean surface after which the extraplacental membranes and umbilical cord were trimmed from the placenta. The fetal surface was photographed with the Lab ID number and 3 cm. of a plastic ruler in the field of view using a standard high-resolution digital camera (minimum image size 2.3 megapixels). A trained observer (D.H.) captured series of *x,y* coordinates that marked the site of the umbilical cord insertion, the perimeter of the fetal surface, and the “vascular end points”, the sites at which the chorionic vessels disappeared from the fetal surface. The perimeter coordinates were captured at intervals of between 1cm and 2cm, and more coordinates were captured if it appeared essential to accurately capturing the shape of the fetal surface.

Placentas were cut perpendicularly to the chorionic plate by a series of parallel cross-sections. The cuts were generally made starting from the apparent geometric center towards the edge of the placental surface creating 8 slices and 7 unique cross-sections. The frequency of the cross-sections was increased if it appeared necessary to capture significant details of placental shape, such as multiple lobes. Each placental slice was digitally traced by a trained observer (D.H.) to capture its perimeter. The perimeter coordinates were again captured at intervals of between 1 and 2cm.

Placentas which were received from the Pathology Department in several pieces, or with parts removed could not be sliced according to the protocol. This has resulted in a reduced number of cut placentas, compared to the number of placentas collected. The total number of placentas for which the traced slices were obtained is 771, of which N=587 were delivered at term. This equals 57% of all placentas delivered at term in the study.

*Fourier analysis of the placental surface shape.*

As in our previous work [2], we use Fourier analysis to quantify the placental surface shape. First, the umbilical insertion point is placed at the origin. Perimeter markers are connected by straight line segments to obtain an approximate perimeter $P$ of the chorionic plate. A sector of opening of 6° with vertex at the origin is rotated in 6° increments. For each turn of the sector, the points in $P$ inside of it are averaged to yield a radial marker. In this way, we obtain 60 radii emanating from the origin spaced at 6° intervals. They are connected to obtain the *angular* *radius* r(θ), which is a function of the angle θ from the umbilical insertion point. For the function r(θ) we compute its Fourier coefficients $C_n$ as

$$C_n = \frac{1}{2\pi}\int_0^{2\pi} e^{-in\theta} r(\theta)\, d\theta$$

Fourier coefficients are a useful tool, in particular, since we would like to identify the typical regularly-irregular placental surface shapes, such as tri-lobate, and star-shaped placentas (see Fig. 1 and Fig 2). An n-fold symmetry of the placental surface is reflected in a larger value of the coefficient $C_n$. In Fig. 3 we demonstrate this principle with two idealized placental shapes: a tri-lobate one ($C_3$=1, and for all other n>0, $C_n$=0), and a star-shaped one ($C_5$=1, and for all other n>0, $C_n$=0). Observe, how this principle works for also for real regularly-irregular placentas in Fig 1.

*Quantifying the displacement of the umbilical cord insertion.*

We have shown in [2] that the centrality of the insertion of the umbilical cord has a significant impact on the placental functional efficiency. As in [2], we use the Fourier coefficient $C_1$ to measure the non-centrality of the insertion, and call it the *Fourier displacement* of the umbilical cord. If the shape is circular, with a central cord insertion, then $C_1$=0. A large Fourier displacement of the umbilical cord insertion does not influence the normal round shape and of a normal diameter. However, a large displacement results in a heavier placenta [2], with a reduced functional efficiency per unit of placental mass: larger β, which means a smaller baby for a given placental weight.

*Quantifying the abnormality of placental surface shape and deformation of placental vascular tree.*

We use Fourier coefficients $C_2$, $C_3$, $C_4$, $C_5$ to describe the variability of the shape of the placental surface. Our DLA (Diffusion Limited Aggregation) stochastic growth model of placental vasculature [1] predicts that a deformed surface shape reflects an abnormal development of the placental vascular tree. Typical deformations of the vascular growth described in [1] result in star-shaped or multi-lobate placentas. Quantitatively, such a placental shape has large values of $C_2$, $C_3$, $C_4$, $C_5$. The size of the coefficient reflects the symmetry of the placental shape; e.g. a tri-lobate placenta will have a larger $C_3$ (see Figure 1).

We also use a measurement of deviation of placental shape from the mean round shape described in [7] as *Symmetric Difference* $\Delta$. The larger value of $\Delta$ implies a larger difference between the placental surface area and the mean round shape.

Our DLA model predicts that the vasculature of a regularly-irregular shaped placenta fills the surface unevenly [1 4], with gaps in vascular coverage forming sectors centered on the umbilical insertion point. This would produce a cross-section of a highly variable thickness, with thinner areas corresponding to the sectors with less vascular coverage.

*Quantifying the placental thickness.*

To measure placental thickness, we analyze the *central* cross-cut of the placenta, which approximates the longer placental diameter. We parameterize the top and bottom boundaries of the central slice by their length. We then put 100 markers on top and bottom of the slice so that the length of the boundary piece between two markers is constant. The markers are grouped into opposite pairs.

We measure the following quantities:

a) *mean placental thickness* is the mean distance between the opposing markers;

b) *normalized mean placental thickness* is the mean placental thickness divided by the width of the slice;

c) *variability of thickness* for each pair of opposite markers $Marker_i^{top}$, $Marker_i^{bottom}$ we calculate $d_i$=distance($Marker_i^{top}$, $Marker_i^{bottom}$ ), which is the thickness of the slice measured between the two markers. We then calculate the absolute value of the deviation of $d_i$ from  from mean placental thickness. We average the deviations over all pairs of markers, and divide the resulting quantity by the value of the mean placental thickness. The result is a number between 0 and 1, with 0 corresponding to no variability (uniform thickness).

*Measuring placental functional efficiency.*

As we have shown in [3,4] the birth weight of a newborn (BW) does not scale linearly with the placental weight (PW). The scaling relation between the two quantities is

$$PW \sim BW^{3/4},$$

which is a version of Kleiber's Law, reflecting the fractal structure of the placental vasculature. We have introduced a measure of the placental functional efficiency

$$\beta = \log PW / \log BW.$$

A larger value of $\beta$ corresponds to a lower birth weight for a given placental weight, and hence a lowered placental function per unit of volume.

*Software.*

Numerical simulations of vascular trees were carried out using our ANSI C, 3-dimensional, diffusion-limited aggregation simulation package "*DLA-3d-placenta*", developed under the terms of the GNU General Public License as published by Free Software Foundation. For DLA cluster visualization we have used *PovRay*: a freeware ray tracing program available for a variety of computer platforms.  Fourier coefficients were calculated using *Maplesoft Maple 12.0* Mathematics and Engineering software. The analysis of the digitally traced placental surfaces and slices was carried out using our proprietory ANSI C package "Placental-geometry".

## Results.

*Thickness versus Fourier displacement.*

The first Fourier coefficient $C_1$ is strongly and significantly *positively* correlated with mean thickness (Pearson's r=0.146, p<0.001; Spearman's ρ=0.128, p=0.002). Thus a large umbilical cord displacement results in a thicker placenta. This confirms the findings of [2].

*Variability of thickness versus variability of the shape.*

Our present measurements support the predictions of the DLA model [1], see Figure 2. The coefficients $C_2$, $C_3$, $C_4$, $C_5$ are strongly and significantly correlated with the variability of thickness (see Table 1). Further, there is a negative correlation with mean thickness, and with normalized mean thickness. Thus, the deformation of the placental surface shape which corresponds to a deformed placental vascular growth, (which may, in particular, be manifested as a regularly-irregular shape), results in a higher variability of placental thickness, and a lower mean thickness.

*Variability of thickness versus placental efficiency.*

Finally, we test whether the effect of the lower placental thickness is a reduced placental efficiency, as predicted by our model [1,4]. We find a strong and significant *positive* correlation of variability of thickness with β. Thus, a placenta with thin parts will tend to have a reduced functional efficiency, resulting in a smaller baby for a given placental weight.

Using the symmetric difference Δ, thickness, and umbilical cord displacement as control variables, we find a significant (p<0.004) partial correlation (0.135) between the value of β and variability of thickness.

**Conclusions.**

Our findings confirm the predictions of the empirical model of placental vascular growth [1,2], as well as demonstrate the effectiveness of Fourier analysis of the placental chorionic surface shape. We find two distinct scenarios in which the morphology of the placental surface affects the placental thickness. In the first, the umbilical cord is inserted non-centrally. As shown in [2], there is no abnormality in the placental surface shape in this case. Yet, there is an increase in the placental weight, relative to the birth weight. This is confirmed by our present results, which show a higher average thickness of the placenta in this case. The placental functional efficiency is lowered, which is reflected in a higher value of the scaling exponent $\beta$.

The second scenario corresponds to a deformed placental surface shape, quantified as a higher value of one of the Fourier coefficients $C_2,..,C_5$. This case includes, in particular, the typical regularly-irregular placental shapes, such as multi-lobate or star-shaped placentas. We have speculated [1] that these deformed shapes reflect irregular villous arborization due to the impact of stressors on the placental vascular branching growth. We have proposed in [1], based on our empirical model of placental vascular growth, that such surface shapes will be accompanied by irregular placental thickness, with thinner areas corresponding to gaps in vascular coverage. This is confirmed by the findings of the present paper. Deformed surface shapes correspond to highly variable thickness. The presence of areas of low thickness leads to a lower average thickness in this case. As expected [1,4], such placentas also have a lowered functional efficiency (a higher value of $\beta$), and thus yield a smaller baby for a given placental weight.

Accounting for effects of placental shape, cord displacement, and the value of thickness as factors, we see that the variability of thickness alone is responsible for much of the observed reduction in placental functional efficiency. Thus, the non-uniform villous and vascular arborization, which is reflected in a non-uniformly thick placenta, is reflected in a lower birth weight for a given placental weight (larger $\beta$).

The measurement of variability of placental thickness is thus shown to be a valuable predictor of placental functional efficiency, reflecting deficiencies of placental vascular development which may not be detectable from the placental surface shape.

Our approach to the study of placental thickness has been motivated by the qualitative modeling of the fractal growth of the placental vasculature started in [1]. Other authors have studied the fractal structure of the placental vascular tree and its impact on the placental function (e.g. [8,9,10]). We expect this approach to yield further insights.

This work was partially supported by NSERC Discovery Grant (M. Yampolsky), by NARSAD Young Investigator Award (C. Salafia), by K23 MidCareer Development Award NIMH K23MH06785 (C. Salafia).

**Figure 1: placental shape and thickness.** (a) tri-lobate placenta: large $C_3$, small mean thickness; (b) bi-lobate placenta: large $C_2$, small mean thickness; (c) normally shaped placenta.

| Traced placental perimeter | Traced central slice | Fourier coefficients | Mean thickness | Normalized mean thickness |
|---|---|---|---|---|
| (a) 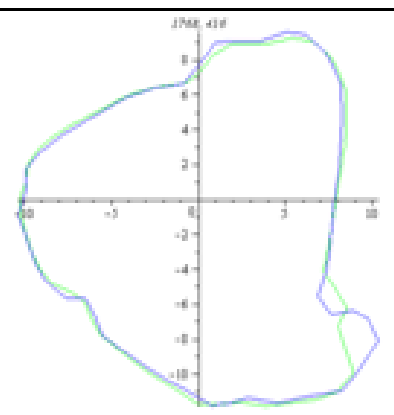 | 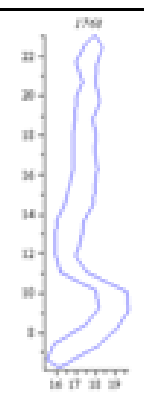 | $C_1$=1.57<br>$C_2$=0.80<br>$C_3$=1.42<br>$C_4$=0.7<br>$C_5$=0.25 | 1.134 | 0.071 |
| (b) 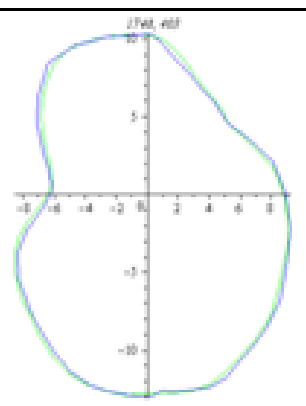 | 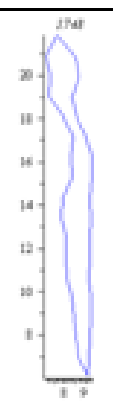 | $C_1$=1.78<br>$C_2$=2.18<br>$C_3$=1.07<br>$C_4$=0.39<br>$C_5$=0.38 | 1.005 | 0.072 |
| (c) 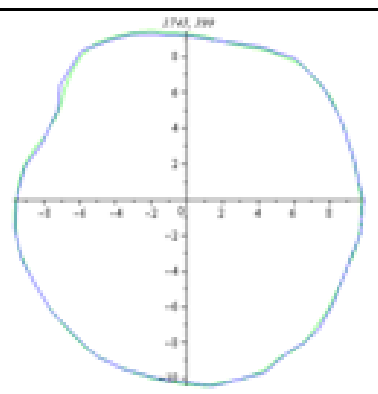 | 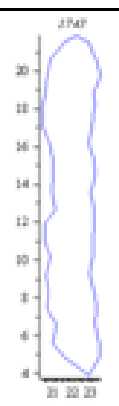 | $C_1$=0.64<br>$C_2$=0.49<br>$C_3$=0.34<br>$C_4$=0.46<br>$C_5$=0.28 | 2.243 | 0.134 |

**Figure 2: variability of thickness for irregularly-shaped placentas described by the DLA model.** Above: the slices of a round placenta have uniform thickness. Below: the slices of a star-shaped placentas have highly variable thickness**.**

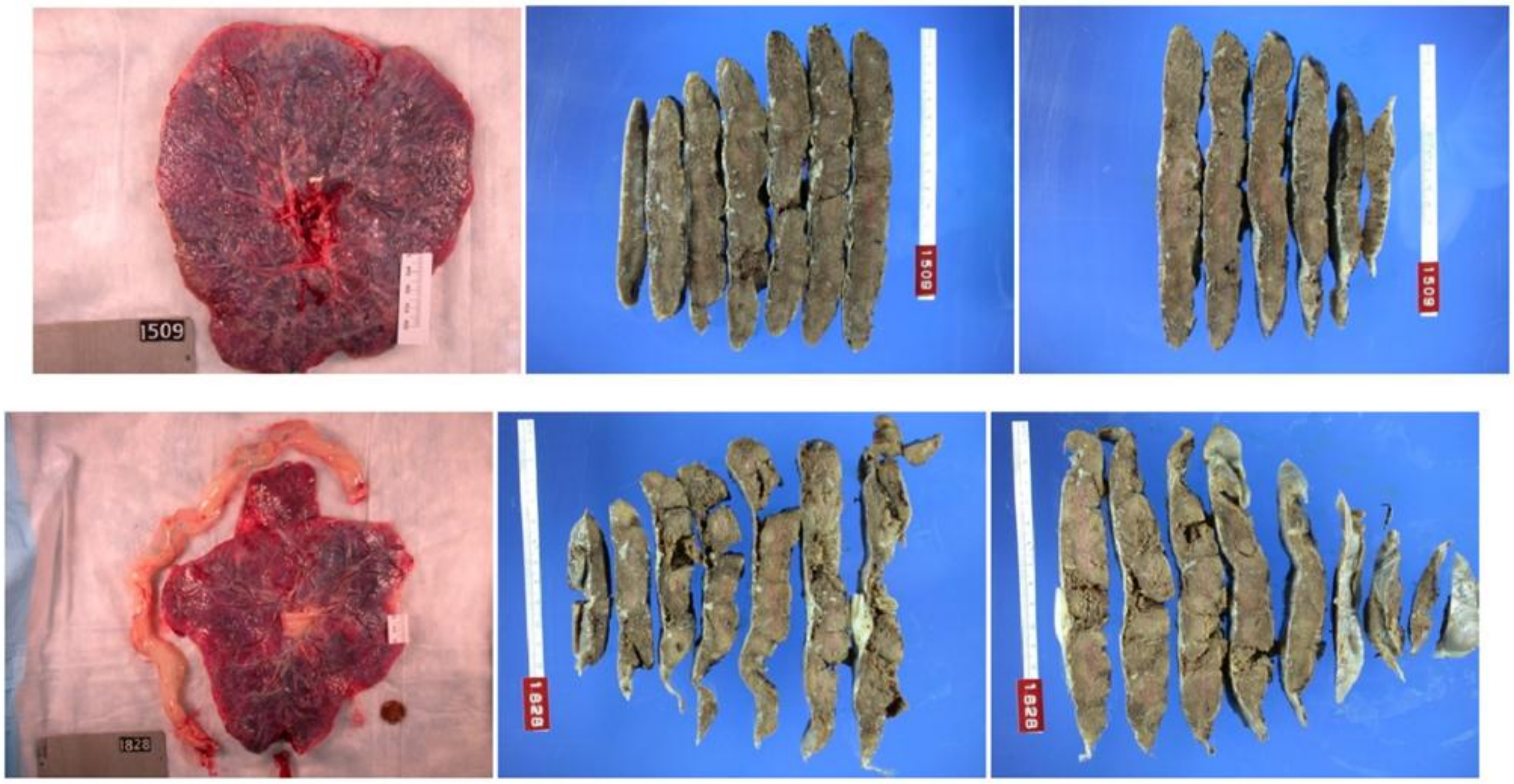

**Figure 3: idealized placental shapes and Fourier coefficients.** Left: idealized tri-lobate placenta, $C_3$=1, for all other n>0, $C_n$=0. Right: idealized star-shaped placenta, $C_5$=1, for all other n>1, $C_n$=0.

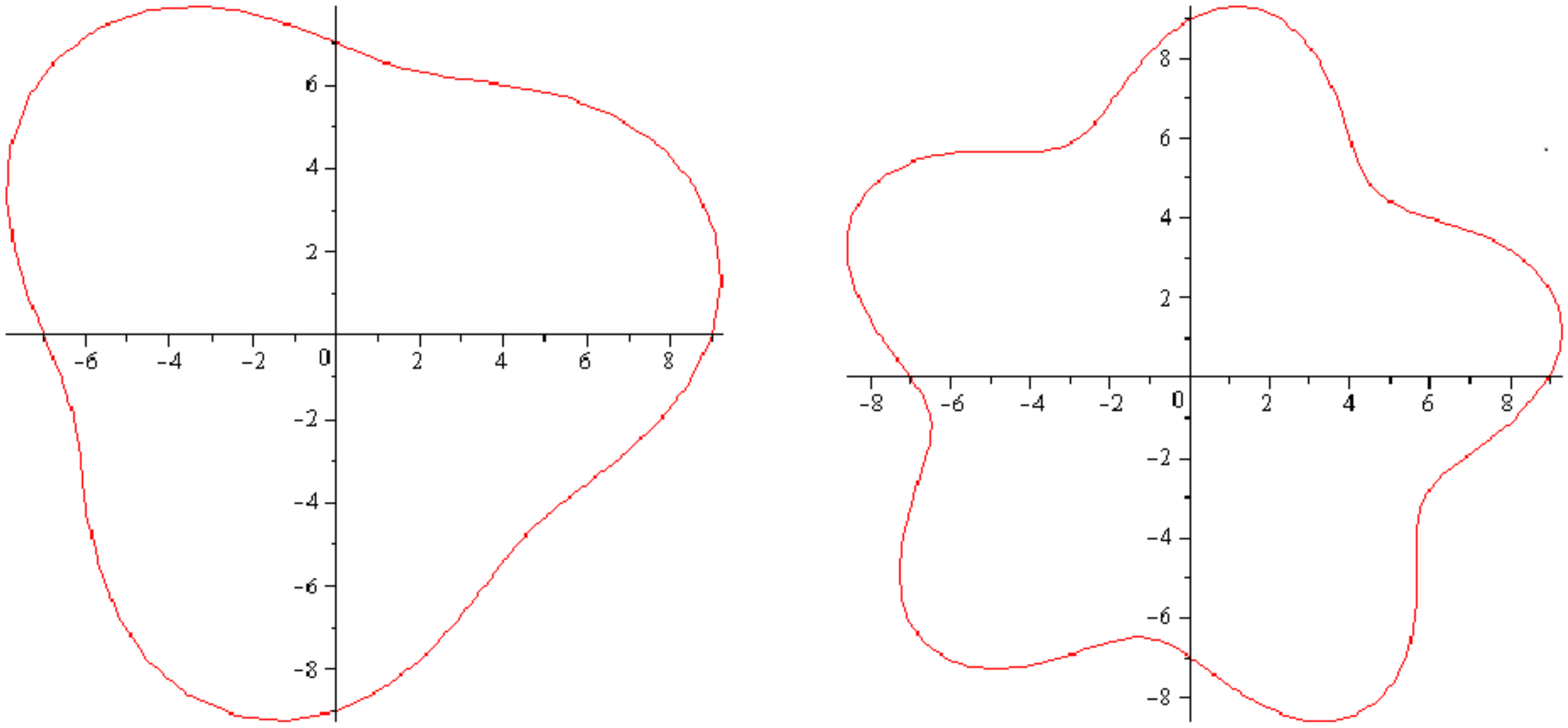

**Table 1. Correlations of Fourier coefficients with thickness variables.**

| | $C_1$ | $C_2$ | $C_3$ | $C_4$ | $C_5$ | β |
|---|---|---|---|---|---|---|
| **Mean thickness** | **r=0.146**<br>**(p<0.001 )**<br>**ρ=0.128**<br>**(p =0.002)** | **r=-0.91**<br>**(p=0.028 )**<br>ρ=-0.76<br>(p=0.66 ) | **r=-0.172**<br>**(p<0.001 )**<br>**ρ=-0.135**<br>**(p=0.001 )** | **r=-0.184**<br>**(p <0.001)**<br>**ρ=-0.185**<br>**(p <0.001)** | **r=-0.145**<br>**(p <0.001)**<br>**ρ=-0.114**<br>**(p =0.006)** | **r=0.454**<br>**(p<0.001 )**<br>**ρ=0.467**<br>**(p <0.001)** |
| **Normalized mean thickness** | r=0.064<br>(p =0.121)<br>ρ=0.060<br>(p =0.147) | **r=-0.189**<br>**(p<0.001 )**<br>**ρ=-0.175**<br>**(p <0.001)** | **r=-0.267**<br>**(p <0.001)**<br>**ρ=-0.214**<br>**(p<0.001 )** | **r=-0.244**<br>**(p <0.001)**<br>**ρ=-0.234**<br>**(p <0.001)** | **r=-0.222**<br>**(p <0.001)**<br>**ρ=-0.173**<br>**(p<0.001 )** | **r=0.210**<br>**(p<0.001 )**<br>**ρ=0.254**<br>**(p<0.001 )** |
| **Variability of thickness** | **r=0.102**<br>**(p=0.013 )**<br>**ρ=0.101**<br>**(p=0.014 )** | **r=0.114**<br>**(p=0.006 )**<br>ρ=0.057<br>(p =0.171) | **r=0.154**<br>**(p <0.001)**<br>**ρ=0.111**<br>**(p=0.007 )** | **r=0.149**<br>**(p<0.001 )**<br>**ρ=0.102**<br>**(p =0.013)** | **r=0.146**<br>**(p<0.001 )**<br>**ρ=0.081**<br>**(p =0.050)** | **r=0.076**<br>**(p =0.065)**<br>**ρ=0.086**<br>**(p=0.038 )** |